\pdfoutput=1
\documentclass[9pt,technote]{IEEEtran}
\ifCLASSINFOpdf
\else
\fi
\usepackage{graphicx}

\usepackage[cmex10]{amsmath}

\begin{document}
%
\title{Synchronization and quorum sensing in  a swarm of humanoid robots}
%
%
%

\author{Patrick Bechon, 
        Jean-Jacques Slotine%
\thanks{Patrick Bechon is a student from Ecole Polyetchnique, France. He is currently studying at MIT in the Nonlinear Systems Laboratory.}
\thanks{Jean-Jacques Slotine heads the Nonlinear Systems Laboratory at MIT, USA}}

\maketitle

\begin{abstract}
With the advent of inexpensive simple humanoid robots, new classes of robotic questions can be considered experimentally. One of these is collective behavior of groups of humanoid robots, and in particular robot synchronization and swarming.  The goal of this work is to robustly synchronize a group of humanoid robots, and to demonstrate the approach experimentally on a choreography of 8 robots. We aim to be robust to network latencies, and to allow robots to join or leave the group at any time (for example a fallen robot should be able to stand up to rejoin the choreography). Contraction theory is used to allow each robot in the group to synchronize to a common virtual oscillator, and quorum sensing strategies are exploited to fit within the available bandwidth.  The humanoids used are Nao's, developed by Aldebaran Robotics.
\end{abstract}

\begin{IEEEkeywords}
Nao, synchronization, contraction, quorum sensing.
\end{IEEEkeywords}

%
\IEEEpeerreviewmaketitle

\section{Introduction}
%
%
%
%
\IEEEPARstart{W}{ith} the advent of inexpensive simple humanoid robots, new classes of robotic questions can be considered experimentally. One of these is collective behavior of groups of humanoid robots, and in particular robot synchronization and swarming. Among the great number of humanoid robot projects, one can think of the first robot to walk Asimo by Honda \cite{sakagami2002intelligent}, of the Humanoid Robotics Projet supported by AIST \cite{akachi2005development} or Darwin created at Virginia Tech \cite{muecke2007darwin}. Toyota has also been working on his partner robots \cite{ota2010partner}. The Robocup (a competition of robot soccer) has also created a league for humanoid robots, and has adopted Nao, a french humanoid robot by Aldebaran Robotics \cite{gouaillier2008nao}, as it standard platform for the Standard Platform League.

This work is the result of a collaboration between MIT's Nonlinear System Laboratory and Aldebaran Robotics, the French company who designed Nao. Nao is used worldwide in universities or in schools, mostly for research or education purposes. 

The aim of this work is to develop a program to synchronize the movements of a large group of robots, for example to realize a synchronized choreography. This project is different from most of the existing work on robot groups because it deals with strong synchronization rather than cooperation, like in Robocup. This kind of experiment has already been conducted by Aldebaran for the Universal Exposition in Shanghai in 2010 with 20 robots. Their approach was to synchronize every robot's clock by using the NTP protocol \cite{mills1991internet}, which is currently use by every computer to synchronize its clock on Internet. Then all robots start the choreography at the same time, assuming every that every robot will have the same loading time. The drawback of this system is that it is not possible to compensate for an small error on one robot (if it starts with a delay for example), or to add a new robot during the choreography. So we tried to develop a more robust way to synchronize this behavior.

Since our goal is really to ensure that every robot is acting precisely with the others (to do synchronization rather than collaboration), the issues that have to be solved are almost entirely network-related. If the robots had access to a perfect network (with immediate transmission of data between all the robots), a master-robot would have been able to sent immediate commands for every other robot, and they would have been synchronized. But with real networks, information takes time to propagate and this time can be highly variable. It can be as small as several milliseconds, but it can also take several seconds in worst case, if the network has too much load for instance, and especially if using Wi-Fi.

To synchronize clocks betweens two robots, we also use the NTP protocol. It allows every robot to share the same time, with an offset of a few milliseconds in worst case. But the robot will constantly adjust his speed to catch up with every other robot.

To achieve this goal our work is based on two theoretical ideas : contraction theory (already used in another case of synchronization between humanoid robots \cite{walkcontractionsync}) and quorum sensing \cite{russo2010global} .

A video illustrating the result of this work is available on the Web, on the Aldebaran Robotics Youtube Channel. \footnote{direct link : http://www.youtube.com/watch?v=emFM8xaQkK4}

\section{Contraction}

Contraction theory provides a framework to guarantee convergence of complex nonlinear systems, without using information concerning the limit trajectory \cite{lohmiller1998contraction}.

\subsection{Contraction definition}

Let consider a system of the form : 

\[ \dot{x} = f(x,t) \]

where $x$ is a size $n$ vector. Let call $\delta x$ the virtual displacement. This virtual displacement is an infinitesimal displacement at fixed time. Its norm is then a measure of the difference between two trajectories. So the evolution of $\delta x(x_0, t)$ is the evolution of the difference between the two trajectories with initial conditions at $t = 0$, $x = x_0$ and $x = x_0 + \delta x(x_0, t = 0)$.

The evolution of $\delta x$ is given by :

\[ \delta \dot{x} = \frac{\partial f}{ \partial x} (x,t) \delta x \]

It is then possible to calculate the evolution of the difference between two trajectories initially close :

\[ \frac{d}{dt} (\delta x ^{T} \delta x) = 2 \delta x ^{T} \delta \dot{x} = 2 \delta x ^{T} \frac{\partial f}{ \partial x} (x,t) \delta x\]

Let $\lambda_{max}$ be the higher eigenvalue of the hermitian part of $\frac{\partial f}{ \partial x}$.

\[ \frac{d}{dt} (\delta x ^{T} \delta x) \le 2 \lambda_{max} \delta x ^{T} \delta x \]

By integration, we can upper bound $\delta x ^{T} \delta x$ as

\[ \| \delta x \| \le \| \delta x_{0} \| e^{\int_{0}^{t} \lambda_{max}(x,t) dt} \]

If $\lambda_{max}$ is negative on the whole trajectory, $\delta x$ will tend toward 0, and this for every initial condition. If $\lambda_{max}$ is negative in a stable region, every system starting in this region will tend toward a limit trajectory, independently of its initial condition. In this case we say that the system is \emph{contracting} in this region. It will forget its initial condition exponentially fast.

\subsection{Generalization}

It's possible to generalize the previous proof by noticing that it does not depend on the definition of length. So we can use a more general definition of the length to prove the contraction definition, the exponential convergence will remain true in any other metric. The details of this generalization can be found in \cite{lohmiller1998contraction}.

We can define this metric change by a matrix $\Theta$, such as new coordinates are of the form $\delta z = \Theta \delta x$. Since this equation is not integrable in general, it's not always possible to define explicitly $z$ in term of $x$. In the new coordinate system, the system is contracting if the largest eigenvalue of the symmetric part of 

\[ F = \left( \dot{\Theta} + \Theta \frac{\partial f}{ \partial x} (x,t) \right) \Theta^{-1} \] 

is negative

\subsection{Andronov-Hopf oscillator}

Consider the classical Andronov-Hopf oscillator, \cite{pham2007stable}.

\begin{equation}
 \left\{ \begin{array}{l}
  \dot{x} + (x^2 + y^2 - 1)x + y = 0 \\
  \dot{y} + (x^2 + y^2 - 1)y - x = 0          
\end{array} \right.
\label{eq_andronov}
\end{equation}

\label{preuve_aondronov}

 Its trajectories converge to the limit cycle $x(t) = sin(t + \theta), y(t) = cos(t + \theta)$ , as can be shown applying the invariant set theorem to the Lyapunov-LaSalle function $E = x^2 + y^2 -1$ \cite{Slotine1991fk} .
 



 
This is an oscillator so obviously it is not contracting. The goal is to find a coupling between each oscillator to have all of them converge toward the same trajectory. Let consider a coupling between every oscillator, proportional to the difference of position in $x$ and $y$.
 
 \[ \left\{ \begin{array}{l}
  \dot{x}_i + (x^2_i + y^2_i - 1)x_i + y_i = \kappa \sum_{j = 0}^{N} (x_j - x_i) \\
  \dot{y}_i + (x^2_i + y^2_i - 1)y_i - x_i = \kappa \sum_{j = 0}^{N} (y_j - y_i)          
\end{array} \right.
 \]
 
The contraction of this system is shown in \cite{pham2007stable}. By adding $N \kappa x_i$ in the first equation and $N \kappa y_i$ in the second, we can have the same right hand side term for every oscillator : 

\[ \left\{ \begin{array}{l}
  \dot{x} + (x^2 + y^2 + \beta)x + y = u(t) \\
  \dot{y} + (x^2 + y^2 + \beta)y - x = u(t)          
\end{array} \right.
 \]

where $\beta = \kappa N - 1$ and $u(t) = \kappa \sum_{j = 0}^{N} x_j$.
 
\[ \frac{\partial f}{\partial x}(x,t) = - \left[ \begin{array}{cc}
	3x^2 + y^2 + \beta & 2yx + 1  \\
	2xy - 1 & x^2 + 3y^2 + \beta
\end{array} \right] \]

The eigenvalues of the symmetric part of this matrix are $\beta + x^2 + y^2$ and $\beta + 3x^2 + 3y^2$. So if $\beta > 0$ the system is contracting, and the contraction rate is at least $\beta$. So the distance between two trajectories will decrease at least like $e^{-\beta t}$ \cite{belabbas2010factorizations, pham2007stable}.

\begin{figure}[!h]
\begin{center}

\includegraphics[width=8cm]{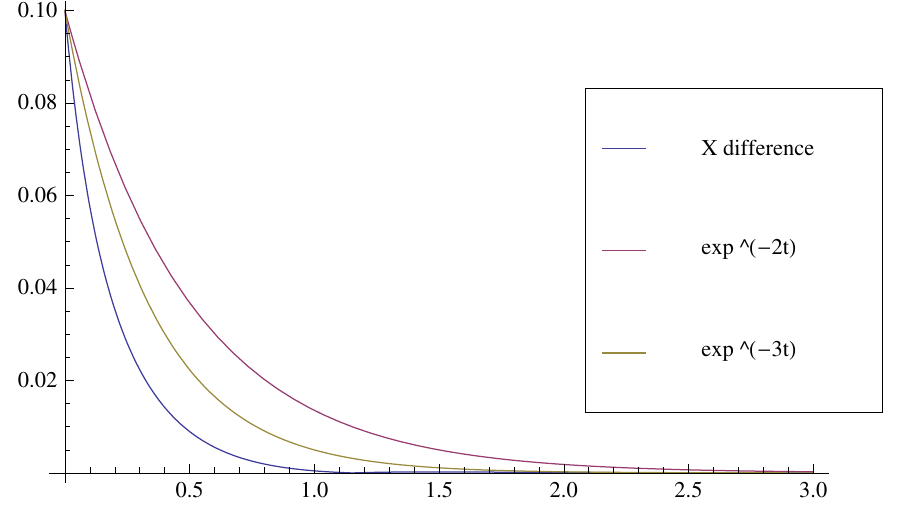}
\caption{ Evolution of $| x_1 - x_3|$ and of the contraction rate}
\label{simulation_x_difference_fig}

\end{center}
\end{figure}

\begin{figure}[!h]
\begin{center}

\includegraphics[width=8cm]{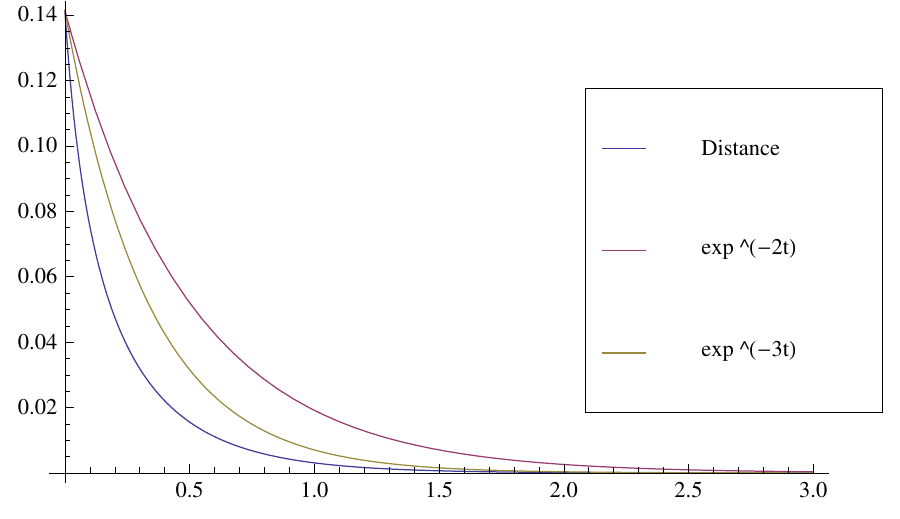}
\caption{ Evolution of $\sqrt{(x_1 - x_3)^2 + (y_1 - y_3)^2}$ and of the contraction rate.}
\label{simulation_distance_difference_fig}

\end{center}
\end{figure}

\begin{figure}[!h]
\begin{center}

\includegraphics[width=8cm]{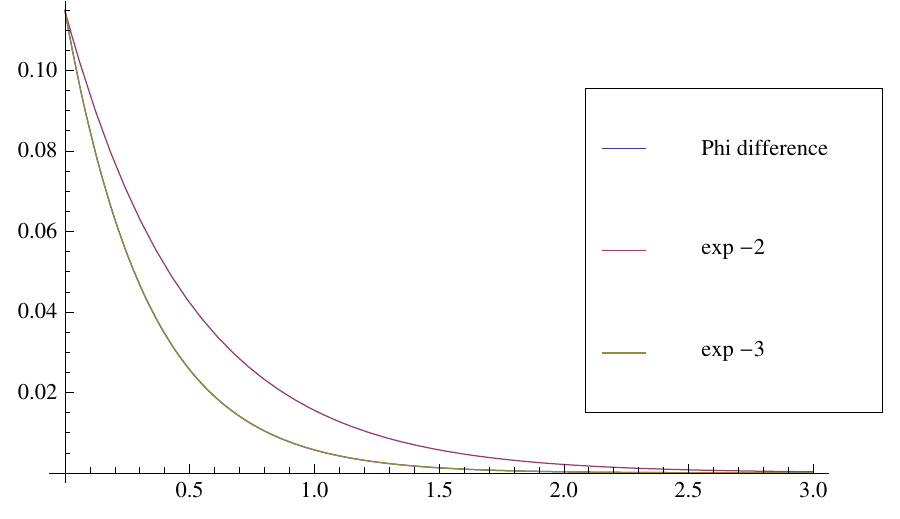}
\caption{ Evolution of $| \phi_1 - \phi_3|$ and of the contraction rate. The brown curve is overlaying the blue one.}
\label{simulation_phi_difference_fig}

\end{center}
\end{figure}

 So it is at least $\beta = N \kappa - 1$, and close to the limit cycle (where $x^2 + y^2 = 1$), it's at least  $\beta = N \kappa$. Here is a simulation of three oscillators, with $\kappa = 1$ and initial conditions $x_i(0) = -1 + \frac{i}{10}$ and $y_i(0)= 2 +\frac{i}{10}$. So the three oscillators starts in neighboring trajectories. And we have plotted on each of the following figure the evolution of a distance between oscillator 1 and 3, and the curves $e^{- (N \kappa - 1) t}$ and  $e^{- (N \kappa) t}$.  Fig. \ref{simulation_x_difference_fig} shows the difference in x,  Fig. \ref{simulation_distance_difference_fig} shows the distance on the phase diagram, and Fig. \ref{simulation_phi_difference_fig} shows the difference in $\phi$.

So every oscillator, independently of its initial conditions, will rejoin the other oscillators and oscillate with them. The difference between two oscillators will decrease with a exponentially speed proportional to $\beta$. The limit cycle of the oscillators will be of the form $x(t) = sin(t + \theta), y(t) = cos(t + \theta)$.

\section{Quorum sensing}

To have the best synchronization possible, it is necessary to have a fully-connected network - that is to say,  every robot must have access to every other robot position, because if the distance between two robots increases, the time taken to detect and correct this difference will also increase.

But on the other hand, the more links we have in the same network the more stress the network will have. So adding links will decrease the overall quality of every link and will also decrease the number of robots that will be able to synchronize in the same time on a given network. To simulate a fully-connected network while limiting the number of links, we can find inspiration in natural mechanisms, such as quorum sensing. The idea is to communicate through a global variable that everyone can access and modify, rather than with every other robot \cite{russo2010global} \cite{tabareau2010noise}. This strategy also reduces considerably the number of necessary connections. In biology~\cite{Bassler}, agents (bacteria) can emit a fixed quantity of chemical (so-called auto-inducer), and also measure the concentration of the chemical in its environment so as to have an estimate of the number of agents present locally. As the bacteria multiply, an infection is launched only once the concentration reaches a certain threshold, i.e., once there are enough other agents present and thus a "quorum" is reached. 

This phenomenon is reproduced here to calculate the mean of every oscillator position. It is then possible to couple oscillators directly with the mean, and the result will be the same in theory.

 \[ \left. \begin{array}{l}
  \dot{x}_i + (x^2_i + y^2_i - 1)x_i + y_i = \kappa \sum_{j = 0}^{N} (x_j - x_i) = \\ \qquad \qquad \qquad \qquad  \qquad \qquad \qquad \qquad N \kappa \left[ \frac{1}{N} ( \sum_{j = 0}^{N} x_j) - x_i \right] \\
  \dot{y}_i + (x^2_i + y^2_i - 1)y_i - x_i = \kappa \sum_{j = 0}^{N} (y_j - y_i) = \\ \qquad \qquad \qquad \qquad  \qquad \qquad \qquad \qquad N \kappa \left[ \frac{1}{N} ( \sum_{j = 0}^{N} y_j) - y_i \right]          
\end{array} \right.
 \]

\begin{figure}[!h]
\begin{center}

\includegraphics[width = 8cm, trim = 0.5cm 5cm 0.5cm 5.5cm, clip]{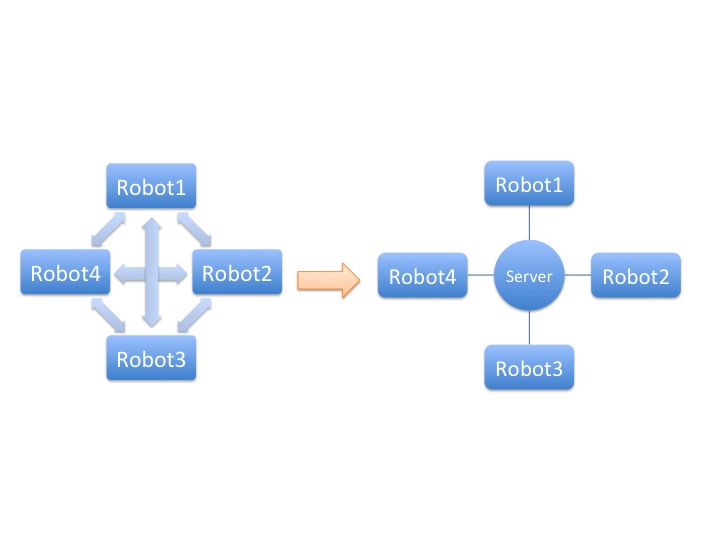}
\caption{Comparaison of the network topology without (left) or with (right) quorum sensing}
\label{comparaison_quorum_sensing}

\end{center}
\end{figure}

To implement this system, it is compulsory to add at least one new node in the network to collect information of position for every node and to send back the mean. So in this case, we will have a star-shaped network (with a number of link proportional to $N$) to have the same synchronization speed of a fully-connected network (with a number of link proportional to $N^2$). See Fig \ref{comparaison_quorum_sensing}

Even if in this simple case there is only one central server to which every robot is connected, this is not a master/slave system. If the server (or the network) has trouble, the only thing the robots will miss will be their synchronization information. The oscillations will continue at the nominal speed, and the synchronization will not occur until the issue is solved. But if they are already synchronized, they will stay synchronized.

It is also possible to easily distribute this system, so as to reduce the load of the server. If several servers are running in the same time, it is only needed that one robot sends its data to several servers to have all the robots on those servers synchronize. In this way, it is possible to have a distributed system if the workload is too large for a single server.

To go further, it is also possible to put this server on every robot, and to use only the first one who join the choreography. If the first robot leaves, the robots have to elect one of their own to run the new server \cite{garcia1982elections}. If the first robot detects a workload too high, it can ask another robot to start its server, which will handle any incoming robot.

 Nevertheless, this system has a drawback : it takes two messages sent on the network to transmit an update from one robot to another. So the effect of any delay on the network will be amplified. To compensate for this, every data sent is sent with its time of emission and its derivative, to be able to predict its new value when needed. This is also used to drop a data that is too old. Alternatively, the same information may be conveyed by a single composite variable combining each data and its derivative \cite{Slotine1991fk}.

\section{Implementation}


In this paper we will focus on moving according to a predefined trajectory, like a choreography.

\subsection{Trajectory description}

We have seen in the two previous sections how to synchronize efficiently oscillators over a network, but we want to synchronize more complex trajectories. Since it is not possible to describe a complex trajectory as the limit cycle of a differential equation, we have to link directly the position of the oscillator to a robot position. We also want the movements to be close enough to nominal movement : the robots can go faster or slower to rejoin the others, but the movement has to be only slightly modified. Once synchronized, the robot has to move according to the nominal movement.

To describe the movement of a robot inside a predefined choreography, one needs only to know its time position inside that choreography. So we only need to convert the position of our oscillator into a parameter going from 0 to 1 uniformly and continuously to cover precisely the initial movement.


\begin{figure}[!h]
\begin{center}

\includegraphics[width=7cm]{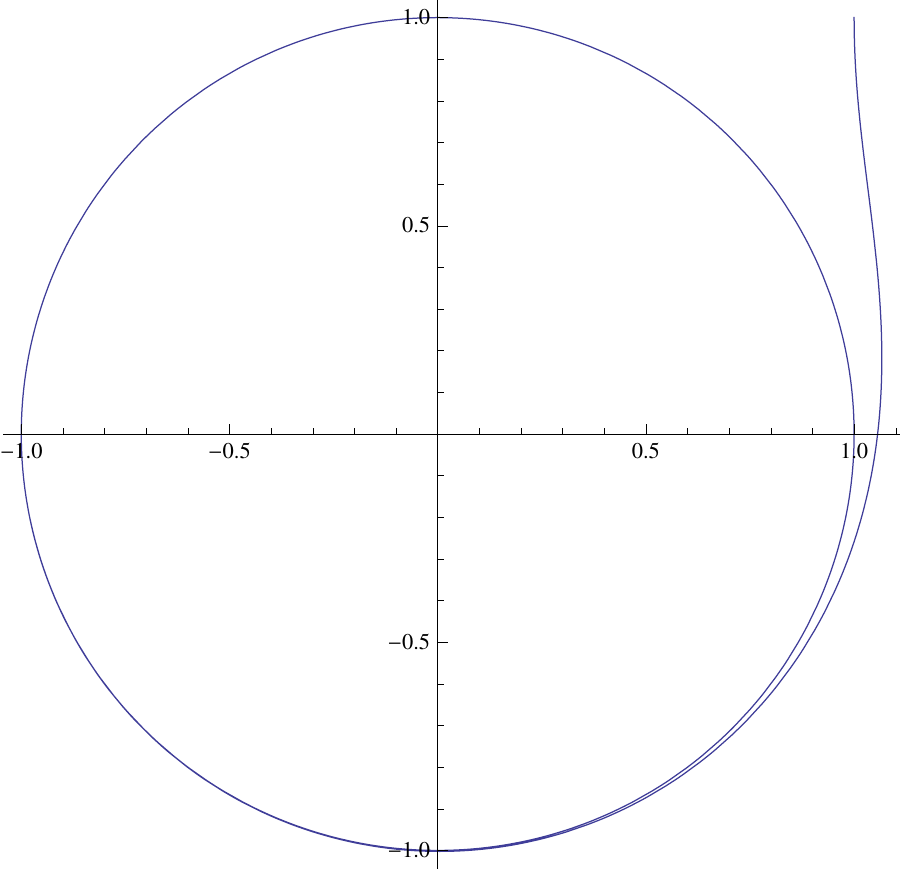}
\caption{Phase diagram of an Andronov-Hopf oscillator (\ref{eq_andronov}) with initial condition x(0) = y(0) = 1.}
\label{portrait_phase_andronov_fig}

\end{center}
\end{figure}

By looking at the phase diagram (Fig. \ref{portrait_phase_andronov_fig}) for an Andronov-Hopf oscillator, we can see that the limit cycle is a perfect circle (and we have proved in section \ref{preuve_aondronov}). By phase diagram we mean here a plot of $x$ in respect of $y$. Indeed when $E$ goes to zero, we have

\[ \left\{ \begin{array}{l}
  \dot{x}  + y = 0 \\
  \dot{y}  - x = 0          
\end{array} \right.
 \]
 
 that is equivalent to $\ddot{y}  + y = 0$. The phase diagram in this case would have been $\dot{y}$ in respect of $y$. We just replace $\dot{y}$ by $x$ even if $E$ is not null for the analogy. 
 
\begin{figure}[!h]
\begin{center}

\includegraphics[width=8cm]{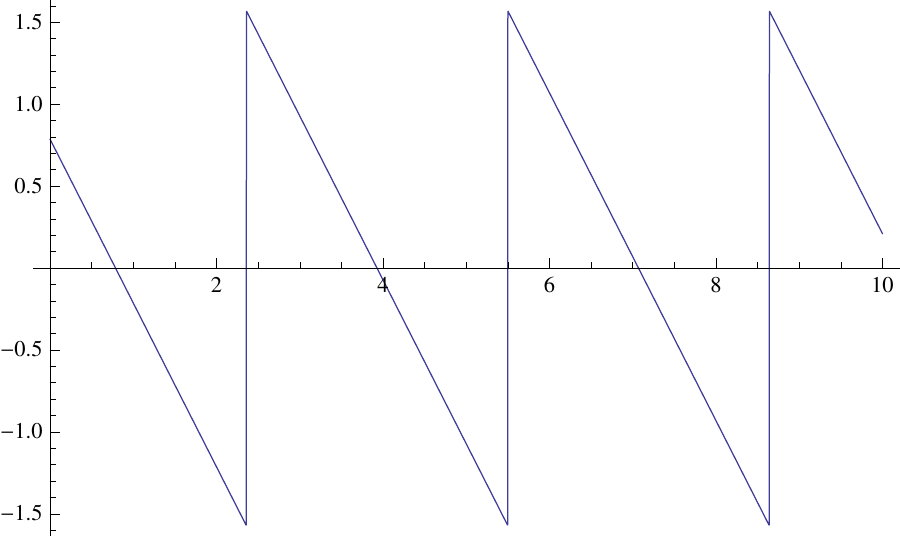}
\caption{ $\phi$ evolution of the Andronov-Hopf oscillator shown in Fig \ref{portrait_phase_andronov_fig}, in radians.}
\label{evolution_phi_andronov_fig}

\end{center}
\end{figure}

 So by considering the angle in this phase portrait, we have a parameter that evolve between 0 and $2\pi$ periodically, and a small variation in the phase diagram (so in the oscillator state) will modify only slightly this angle. The evolution of this angle (called $\phi$) is plotted Fig \ref{evolution_phi_andronov_fig}.

\subsection{Oscillator speed ajustment }

To correctly reproduce a trajectory, it is necessary to adapt to its length, so to adapt the speed of the oscillator. Let $\{x(t), y(t)\}$ be a solution of (\ref{eq_andronov}), and then we will look for a differential equation satisfied by $\{x(\omega t), y(\omega t)\}$. We have $\frac{d}{dt} (y(\omega t)) = \omega \dot{y}(\omega t)$ and $\frac{d}{dt} (x(\omega t)) = \omega \dot{x}(\omega t)$.

\[
 \left\{ \begin{array}{l}
  \omega \dot{x}(\omega t) + (x(\omega t)^2 + y(\omega t)^2 - 1)x(\omega t) + y(\omega t) = 0 \\
  \omega \dot{y}(\omega t) + (x(\omega t)^2 + y(\omega t)^2 - 1)y(\omega t) - x(\omega t) = 0          
\end{array} \right.
\]

So $\{x(\omega t), y(\omega t)\}$ satisfy the differential equation : 

\[
 \left\{ \begin{array}{l}
  \dot{x} + \omega [ (x^2 + y^2 - 1)x + y ] = 0 \\
  \dot{y} + \omega [ (x^2 + y^2 - 1)y - x ] = 0          
\end{array} \right.
\]

By modifying $\omega$ in this equation, we will modify the speed of the oscillator without modifying the trajectory. The phase diagram will stay the same, it will just be covered in a different time. To be precise, the limit cycle will be covered with a period of $\frac{2\pi}{\omega}$. 

To keep the same influence of synchronization, it is also compulsory to replace $\kappa$ by $\omega \kappa$ in previous equation. So the final equation for the oscillator number $i$ is : 

 \[ \left\{ \begin{array}{l}
  \dot{x}_i + \omega [ (x^2_i + y^2_i - 1)x_i + y_i ] = N \omega \kappa \left[ \frac{1}{N} ( \sum_{j = 0}^{N} x_j) - x_i \right] \\
  \dot{y}_i + \omega [ (x^2_i + y^2_i - 1)y_i - x_i  ] = N \omega \kappa \left[ \frac{1}{N} ( \sum_{j = 0}^{N} y_j) - y_i \right]          
\end{array} \right.
 \]

\subsection{Music}

Another issue that has to be considered is the music. We aim at having something to do choreographies, so we need to be able to play music during the performance, and to synchronize the robots with the music. But it is not a good idea to change the speed of the music in real time : in human shows, the music stays at the same pace and only the dancers have to change their pace to synchronize. So we want the music to act as a leader : to share its oscillator position to allow the others to synchronize to itself but not to take into account the other robot positions.

To prove that this behavior will act as we want, we have to use partial contraction \cite{wang2005partial}. The idea is to build a virtual system from the real system, and to prove the contraction of this virtual system to get information on the real system. By proving the contraction of the whole system, we prove that every trajectory will converge to the other. Since the leader trajectory is a particular solution, every trajectory will tend toward this trajectory. Formally, if 

\[\dot{x} = g(x,x,t)\]

is a system, one can build the virtual system

\[\dot{y} = g(y,x,t)\] 

and proving the contraction of this new system will demonstrate that every solution will tend toward $x$ (because $x$ is a particular solution). This theory is developed more in depth in \cite{wang2005partial}. Let $X$ be $[ x, y ]^T$. The system with a leader is : 

\[
\left\{
\begin{array}{l}
\dot{X}_{leader} = f(X_{leader}) \\
\dot{X}_i = f(X_i) +  \kappa \left( N X_i - \sum^N_{j = 1} X_j \right) + \kappa (X_i - X_{leader}), i = 1  ... N
\end{array}
\right.
\]

\[
\left\{
\begin{array}{l}
\dot{X}_{leader} = f(X_{leader}) \\
\dot{X}_i = \underbrace{f(X_i) + \kappa (N+1)X_i}_{\mathrm{= g(X_i), contractant}} \underbrace{ - \kappa \sum^N_{j = 1} X_j - \kappa X_{leader}}_{= u(t)}, \mathrm{for}\  i = 1  ... N
\end{array}
\right.
\]

Here we can recognize in $g(X_i)$ the same system than before : the system which we have already proved the contraction. And $u(t)$ is the same for any robot, so will not have any effects on the synchronization. Thus we have proved the partial contraction of the new system with the leader.

We have then proved that with one robot who share its position without taking into account others' position, the synchronization remains. So this robot can be the musician, and if he start the music with its oscillator, the two will not change their respective speed so they will stay synchronized. This idea can also be used if we want to predict with a good accuracy the time when every robot will finish : they will all finish with the leader, which has a fixed speed.

\subsection{Robot implementation}

\begin{figure}[!h]
\begin{center}

\includegraphics[width=8cm]{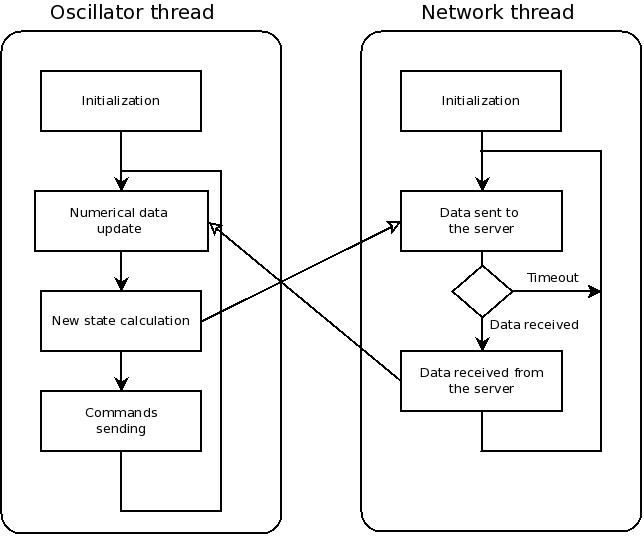}
\caption{Schematic robot implementation of this behavior}
\label{implementation_robot_fig}

\end{center}
\end{figure}

To achieve our initial goal, our implementation has to separate the calculation of the oscillator and the network communication because we don't want any perturbation from the network (latencies, lost message, etc ...) causing a shacking movement on the robot. So we separate the implementation in two different threads : one for the oscillator simulation and joint command and the other one for network communication. A schematic view is shown Fig. \ref{implementation_robot_fig}.

To describe more in detail each task :

\begin{itemize}
\itemsep 4pt
\item \emph{Oscillator initialization} : For this initialization the robot has to determine its initial condition. For this he retrieve data from the server concerning the mean position of the group and estimate the position of group in 1 second. He then has 1 second to join this position, so when he will start to synchronize he will be close enough to the others.
At the end of this step, the network thread can start since there is a virtual oscillator position to be shared with the others.
\item \emph{Numerical data update} : To keep the oscillator calculation real-time, we use a calculation step (used in the numerical resolution of the oscillator differential equation) variable, and equal to the time difference between the beginning of the last loop and the beginning of the current loop. During this tack, the robot calculate this step and use the current time to estimate the current mean of every other robot position (or to discard this information if it's too old, see next section for more information).
\item \emph{New state calculation} : Here the robot can use a Euler method with the current state and the synchronization information estimated on the previous task to calculate the next state of its virtual oscillator. These data are stored in a global variable to be accessed by the network thread.
\item \emph{Commands sending} : With the new state available, a new $\phi$ can be computed and a new robot position too. This new command is then sent to the robot to be reached in 150 ms (in addition to every remaining command). The movement will then be smooth (without interruption) if the calculation step remains under 150ms.
\item \emph{Network thread initialization} : The robot connects itself to the server and request an update of the synchronization information.
\item \emph{Data sent to server} : Sending syntonization data to the server, according to the new section
\item \emph{Data received from server} : Receiving syntonization data to the server, according to the new section

\end{itemize}

\subsection{Server implementation}

The precise goal of this program is to receive the virtual oscillator position from every robot, and to send back to each of them the mean. The server can also receive configuration messages (like changing the coupling strength that is send back to every robot, stopping the server, ...). Synchronization data are transmitted with the form : 

\[
\begin{array}{|c|c|}
\hline
$Message for the server$ & time_1:x:\dot{x}:y:\dot{y}; \\
\hline
$Answer from the serveur$ & time_2:m_x:\dot{m}_x:m_y:\dot{m}_y:N:\kappa;\\
\hline
\end{array}
\]

where $time_1$ is not the sending time, but the time of the calculation that leads to $x$ and $y$. The values of $\dot{x}$ and $\dot{y}$ are send only to do prediction on $x$ and $y$ and are not directly used in the synchronization process. Every value is separated by a '$:$', and every message finishes with a '$;$'. This way, it is easy to parse the incoming messages and separate them.

Every robot expects a message of the form presented above, where $m_x$ is the mean of $x$ and $m_y$ is the mean of $y$. Their derivatives are there for prediction, $N$ is the total number of robot and $\kappa$ is the coupling strength. The server will send a copy of this message to every robot that are connected. Here again, $time_2$ is the time used to evaluate the mean, not the time of the sending. To be more precise $m_x$ is the mean of $x_i + (time_2 - time_1(i))*\dot{x}_i$ and $\dot{m}_x$ is the mean of $\dot{x}_i$ ($i$ is there the robot index). The fact that the server sends $\kappa$ allows the user to change the coupling in real time, so as to activate, increase or decrease its strength.

\section{Perspective of improvements}

In this implementation, the movements of the robots are not really related to the virtual oscillator movements. But for movement like walking, it is possible to do a stronger link and to deal for example with every initial condition \cite{ walkcontractionsync}.

To go further, it could be useful to use the same quorum server to synchronize different types of oscillators. For example if several robots are running different choreography (with different length) or are walking with different pace (because of size differences between robots for example), each robot will have to decompose the quorum signal to find the part of the mean concerning the robot identical to itself. For example if 3 robots walking with 1 step/second and 3 robots with 0.5 step/second, each robot can filter the quorum signal for its corresponding frequency. With a fixed calculation step, it is possible to design filter with an arbitrary precision \cite{Oppenheindiscrete}. With a non-constant step, one has to incorporate the step into the filter coefficients. But without knowing the total number of robots (but with the knowledge of the number of similar robot), a robot can synchronize with all the robots sharing the same speed.

Another improvement could be to track the tempo of the music \cite{goto1995real}, and see it as an oscillator. At the beginning the robot has to determine the frequency to configure its own oscillator, but will then be able to use the music instead of the quorum signal to synchronize. So every robot will synchronize its virtual oscillator to the music, so the movements performed will be synchronized with the music. It can also be a good way for the robot to improvise a dance on a music.

\section*{Acknowledgment}

The authors would like to thank Aldebaran Robotics, and in particular Rodolphe Gelin, for their help in using Nao and their lending of robots for the final video.

\ifCLASSOPTIONcaptionsoff
  \newpage
\fi



%

\bibliographystyle{IEEEtran}
\bibliography{../biblio}

\begin{thebibliography}{10}
\providecommand{\url}[1]{#1}
\csname url@samestyle\endcsname
\providecommand{\newblock}{\relax}
\providecommand{\bibinfo}[2]{#2}
\providecommand{\BIBentrySTDinterwordspacing}{\spaceskip=0pt\relax}
\providecommand{\BIBentryALTinterwordstretchfactor}{4}
\providecommand{\BIBentryALTinterwordspacing}{\spaceskip=\fontdimen2\font plus
\BIBentryALTinterwordstretchfactor\fontdimen3\font minus
  \fontdimen4\font\relax}
\providecommand{\BIBforeignlanguage}[2]{{%
\expandafter\ifx\csname l@#1\endcsname\relax
\typeout{** WARNING: IEEEtran.bst: No hyphenation pattern has been}%
\typeout{** loaded for the language `#1'. Using the pattern for}%
\typeout{** the default language instead.}%
\else
\language=\csname l@#1\endcsname
\fi
#2}}
\providecommand{\BIBdecl}{\relax}
\BIBdecl

\bibitem{sakagami2002intelligent}
Y.~Sakagami, R.~Watanabe, C.~Aoyama, S.~Matsunaga, N.~Higaki, and K.~Fujimura,
  ``The intelligent asimo: System overview and integration,'' in
  \emph{Intelligent Robots and Systems, 2002. IEEE/RSJ International Conference
  on}, vol.~3.\hskip 1em plus 0.5em minus 0.4em\relax Ieee, 2002, pp.
  2478--2483.

\bibitem{akachi2005development}
K.~Akachi, K.~Kaneko, N.~Kanehira, S.~Ota, G.~Miyamori, M.~Hirata, S.~Kajita,
  and F.~Kanehiro, ``Development of humanoid robot hrp-3p,'' in \emph{Humanoid
  Robots, 2005 5th IEEE-RAS International Conference on}.\hskip 1em plus 0.5em
  minus 0.4em\relax IEEE, 2005, pp. 50--55.

\bibitem{muecke2007darwin}
K.~Muecke and D.~Hong, ``Darwin's evolution: development of a humanoid robot,''
  in \emph{Intelligent Robots and Systems, 2007. IROS 2007. IEEE/RSJ
  International Conference on}.\hskip 1em plus 0.5em minus 0.4em\relax IEEE,
  2007, pp. 2574--2575.

\bibitem{ota2010partner}
Y.~Ota, ``Partner robots---from development to implementation,'' in \emph{Human
  System Interactions (HSI), 2010 3rd Conference on}.\hskip 1em plus 0.5em
  minus 0.4em\relax IEEE, 2010, pp. 14--16.

\bibitem{gouaillier2008nao}
D.~Gouaillier, V.~Hugel, P.~Blazevic, C.~Kilner, J.~Monceaux, P.~Lafourcade,
  B.~Marnier, J.~Serre, and B.~Maisonnier, ``The nao humanoid: a combination of
  performance and affordability,'' \emph{CoRR, vol. abs/0807.3223}, 2008.

\bibitem{mills1991internet}
D.~Mills, ``Internet time synchronization: The network time protocol,''
  \emph{Communications, IEEE Transactions on}, vol.~39, no.~10, pp. 1482--1493,
  1991.

\bibitem{walkcontractionsync}
A.~Mukovskiy, J.~Slotine, and M.~Giese, ``Design of the dynamic stability
  properties of the collective behavior of articulated bipeds,'' \emph{IEEE-RAS
  International Conference on Humanoid Robots}, 2010.

\bibitem{russo2010global}
G.~Russo and J.~Slotine, ``Global convergence of quorum-sensing networks,''
  \emph{Physical Review E}, vol.~82, no.~4, p. 041919, 2010.

\bibitem{lohmiller1998contraction}
W.~Lohmiller and J.~Slotine, ``On contraction analysis for non-linear
  systems,'' \emph{Automatica}, vol.~34, no.~6, pp. 683--696, 1998.

\bibitem{pham2007stable}
Q.~Pham and J.~Slotine, ``Stable concurrent synchronization in dynamic system
  networks,'' \emph{Neural Networks}, vol.~20, no.~1, pp. 62--77, 2007.

\bibitem{Slotine1991fk}
J.~Slotine and W.~Li, \emph{Applied Nonlinear Control}.\hskip 1em plus 0.5em
  minus 0.4em\relax Prentice-Hall, 1991.

\bibitem{belabbas2010factorizations}
M.~Belabbas and J.~Slotine, ``Factorizations and partial contraction of
  nonlinear systems,'' in \emph{American Control Conference (ACC), 2010}.\hskip
  1em plus 0.5em minus 0.4em\relax IEEE, 2010, pp. 3440--3445.

\bibitem{tabareau2010noise}
N.~Tabareau, J.~Slotine, and Q.~Pham, ``How synchronization protects from
  noise,'' \emph{PLoS Computational Biology}, vol.~6, no.~1, 2010.

\bibitem{Bassler}
W.~Ng and B.~Bassler, ``Bacterial quorum-sensing network architectures,''
  \emph{Annual review of genetics}, vol.~43, pp. 197--222, 2009.

\bibitem{garcia1982elections}
H.~Garcia-Molina, ``Elections in a distributed computing system,''
  \emph{Computers, IEEE Transactions on}, vol. 100, no.~1, pp. 48--59, 1982.

\bibitem{wang2005partial}
W.~Wang and J.~Slotine, ``On partial contraction analysis for coupled nonlinear
  oscillators,'' \emph{Biological Cybernetics}, vol.~92, no.~1, pp. 38--53,
  2005.

\bibitem{Oppenheindiscrete}
A.~v.~Oppenheim, R.~Schafer, and J.~Buck, \emph{Discrete-time signal
  processing}.\hskip 1em plus 0.5em minus 0.4em\relax Prentice Hall, 1989.

\bibitem{goto1995real}
M.~Goto and Y.~Muraoka, ``A real-time beat tracking system for audio signals,''
  in \emph{Proceedings of the International Computer Music Conference}.\hskip
  1em plus 0.5em minus 0.4em\relax San Francisco: International Computer Music
  Association, 1995, pp. 171--174.

\end{thebibliography}

%




\end{document}